\documentclass{article}

\PassOptionsToPackage{numbers, compress}{natbib}
\usepackage{natbib}

\usepackage[preprint]{neurips_2023}

\usepackage[utf8]{inputenc} 
\usepackage[T1]{fontenc}    
\usepackage[hidelinks]{hyperref}
\usepackage{url}            
\usepackage{booktabs}       
\usepackage{amsfonts}       
\usepackage{nicefrac}       
\usepackage{microtype}      
\usepackage{xcolor}         
\usepackage{amsmath,amssymb}
\usepackage{mathrsfs}

\title{HCC Is All You Need: Alignment---The Sensible Kind Anyway---Is Just Human-Centered Computing}

\author{%
  Eric Gilbert \\
  School of Information \& CSE \\
  University of Michigan \\
  \texttt{eegg@umich.edu} \\
}

\begin{document}

\maketitle

\begin{abstract}
  This article argues that AI Alignment is a type of Human-Centered Computing.
\end{abstract}

\section{Argument}

The argument of this very short paper is that the problem academic \textsc{AI} has termed ``alignment'' is just a type of Human-Centered Computing (\textsc{HCC}). \textsc{HCC} is an existing academic field.\footnote{HCC contains Human-Computer Interaction (\textsc{HCI}).}

The term ``alignment'' rose to prominence among \textsc{AGI} and crypto researchers/enthusiasts/grifters \cite{Ahmed2024}---and has been linked with eugenics traditions \cite{Gebru_Torres_2024}. Nevertheless, it has jumped into academic discourse and become an umbrella term for creating AI systems that respect ``human intentions and values'' \cite{ji2023ai}. I address this latter, arguably more sensible kind of alignment currently happening in academia and industry. Here, alignment often quickly becomes specific technical problems, such as how to learn reward functions that correspond to what users want (e.g., \cite{leike2018scalable}), or how to construct models that can explain themselves to people (e.g., \cite{lipton2018mythos}). However, the high-level goal is broader: to bring a  complex technology into concert with what people want it to do.

This is literally what \textsc{HCC} is. There are whole journals and conferences (e.g., \textsc{CHI}, \textsc{CSCW}, \textsc{UIST}, FAccT). There is a whole program at the U.S. National Science Foundation \cite{hccnsf}.

For over 40 years, \textsc{HCC} has struggled with, and made progress on, ``aligning'' different technologies to people. Key issues include: Who, exactly, are we talking about (e.g., \cite{ackerman2000intellectual,bodker2006second,card2018psychology,eslami2015always,kaur2022sensible,pruitt2003personas,thomas1989minimizing})? How do we know what they want (e.g., \cite{dourish2006implications,friedman2019value,gaver1999design,hayes2011relationship,kraut2004psychological,olson2014ways,terry2023ai})? How stable is what they want to do with technology (e.g., \cite{suchman1987plans})? Can they help us design it (e.g., \cite{haimson2020designing,harrington2021eliciting,kensing1998participatory,madaio2020co,muller1993participatory})? What are the different ways a technology can be designed (e.g., \cite{card1990design,morris2023design})? How do we know if it's good for people (e.g., \cite{ellison2007benefits,meier2024beyond})? What are the limits of design- and tech-centric approaches (e.g., \cite{gansky2022counterfacctual,green2021contestation,neff2020bad,young2022confronting})? Is it possible to avoid baking systemic oppression into technology (e.g., \cite{bardzell2010feminist,dimond2011domestic,erete2018intersectional,friedman1996bias,im2021yes,lam2023sociotechnical,liang2021embracing,ogbonnaya2020critical,sultana2018design})?

Casting alignment as \textsc{HCC} invites it to draw upon the considerable theories, methods, and findings of \textsc{HCC}, and the fields from which it borrows (e.g., \textsc{STS}, Communication, Ethics, etc.)---instead of re-inventing them under new names. Perhaps we don't need the word ``alignment'' at all. 

\section{Argument as equation}
In an effort to observe norms, I have also stated the argument above as the following equation:

\begin{equation}
    \{(AI, L) : L \in \mathscr{L}, AI \in \underset{AI'}{\arg\min}\int_{\mathscr{H}} L(AI',h)\,dh\}  \subset HCC
\end{equation}

where $\mathscr{L}$ represents a family of human-AI loss functions, and $\mathscr{H}$ is the space of what people want.

\pagebreak

\bibliographystyle{ACM-Reference-Format}
\bibliography{refs}

\end{document}